\documentclass[twocolumn,10pt]{svjour3}
\smartqed

\usepackage{graphicx}
\usepackage[update,prepend]{epstopdf}
\usepackage{float} 
\usepackage{bm}
\usepackage{mathptmx}
\usepackage{rotating}
\usepackage{lineno}
\usepackage{txfonts} 
\usepackage{url}
\usepackage[ansinew]{inputenc}
\usepackage{textcomp}
\usepackage{dingbat}
\usepackage[square]{natbib} 

\providecommand{\url}[1]{{#1}}
\providecommand{\urlprefix}{URL }
\expandafter\ifx\csname urlstyle\endcsname\relax
  \providecommand{\doi}[1]{DOI \discretionary{}{}{}#1}\else
  \providecommand{\doi}{DOI \discretionary{}{}{}\begingroup
  \urlstyle{rm}\Url}\fi

\def\be{\begin{equation}}
\def\ee{\end{equation}}
\def\ba{\begin{eqnarray}}
\def\ea{\end{eqnarray}}

\def\g{\gamma}     
\def\d{\delta}     \def\D{\Delta}

\def\la{\label}

\def\lt{\left}
\def\ri{\right}

\newcommand{\grl}{{\it Gepohys. Res. Lett.}}

\newcommand{\sovast}{{\it Soviet Astronomy}}

\journalname{Journal of Geodesy}
\begin{document}
\title{The Fresnel-Fizeau effect and the atmospheric time delay in geodetic VLBI.}
\authorrunning{Kopeikin S.M. \& Han W.-B.}
\titlerunning{Fresnel-Fizeau effect in VLBI time delay}

\author{S.~M. Kopeikin \and W.-B. Han}
\institute{S.~M. Kopeikin \at
              Siberian State Academy of Geodesy, 10 Plakhotny St., Novosibirsk 630108, Russia \\
             \emph{Permanent address: Department of Physics \& Astronomy, University of Missouri, 322 Physics Bldg., Columbia, MO 65211, USA} \\  %  if needed
             \email{kopeikins@missouri.edu} 
           \and
           W.-B. Han \at
              Department of Physics \& Astronomy, University of Missouri, 322 Physics Bldg., Columbia, MO 65211, USA\\
\emph{Permanent address: Shanghai Astronomical Observatory, 80 Nandan Road, Shanghai, 200030, China}
\\
\email{wbhan@shao.ac.cn}
}

\date{Received:     /     Accepted:}

\maketitle

\label{firstpage}

\begin{abstract}
The Fresnel-Fizeau effect is a special relativistic effect that makes the speed of light dependent on the velocity of a transparent, moving medium. We present a theoretical formalism for discussing propagation of electromagnetic signals through the moving Earth atmosphere with taking into account the Fresnel-Fizeau effect. It provides the rigorous relativistic derivation of the atmospheric time delay equation in the consensus model of geodetic VLBI observations which was never published before.  The paper confirms the atmospheric time delay of the consensus VLBI model used in IERS Standards, and provides a firm theoretical basis for calculation of even more subtle relativistic corrections.

\end{abstract}


\section{Introduction}\la{sec1}

Very long baseline interferometry (VLBI) and satellite laser ranging (SLR) are the primary geodetic techniques which are used to measure continental and intercontinental baselines with unprecedented precision \citep{Sovers_1998,Schuh_2012,Collilieux_2007}. The estimated site coordinates for each technique define an international terrestrial reference frame (ITRF) \citep[Chapter 4]{IERS_2010}. Currently, ITRF-VLBI and ITRF-SLR are compared by making use of the standard Helmert (seven-parameter) transformation which provides the translational, rotational, and scale differences between the frames \citep{Torge_2012_book}. Continuous inter-comparison between SLR and VLBI frames consistently indicates to the persisting scale offset between the two realizations of ITRF amounting to 1 ppb, i.e. 6 mm at the Earth's surface \citep{Altamimi_2011,Schuh_2012}. This problem has raised the question whether the relativistic 'consensus' model used by the international Earth rotation service (IERS) for VLBI data analysis \citep{IERS_2010} is entirely self-consistent. 

It motivated a working group of the IAU Commission 52 "Relativity in fundamental astronomy" to start re-examination and improvement of the consensus model in several directions regarding a more appropriate modeling of the VLBI time delay equation in the part concerning the gravitational effects \citep{Kopeikin_2011_book,Soffel_2014}. However, relativity also couples with the propagation of light through terrestrial atmosphere causing a tiny but fundamentally important contribution to the atmospheric time delay. Relativistic equation for the atmospheric time delay was introduced to the consensus model by M. Eubanks on a pure phenomenological basis \citep{eubanks_1991} but it lacked a solid theoretical justification and was never properly modeled from scratch starting from the fundamental VLBI time delay equation applied in the barycentric celestial reference system (BCRS). The present paper is to fulfill this important task.

At first glance the idea to consider the atmospheric delay in the global barycentric coordinates may look strange for the atmospheric delay is related to the local physics of the Earth. It is similarly "local" as, e.g., the tectonic activity or ocean loading. Moreover, the atmospheric delay model (e.g. the Saastamoinen zenith delay and all mapping functions) are defined in a non-relativistic way. This is usually interpreted as a relativistically meaningful formulation in the GCRS \citep{IERS_2010}. For this reason, it seems natural to consider the tropospheric delay as an additional effect applied to the topocentric moments of reception at VLBI stations. In this way, the only relativistic effect which one may expect, would be associated with the relative velocity of one VLBI station with respect to another. Indeed, this is exactly what \citet{eubanks_1991} actually assumed in his consensus model of the relativistic correction to the atmospheric delay that appears as the last term in equation (\ref{1}) below. 

The problem with the naive "geocentric derivation" of the atmospheric delay is that it is completely detached 1) from the propagation model of light in vacuum, and 2) from the set of space-time transformations adopted in the consensus model \citep{Kopeikin_2011_book,Soffel_2014,IERS_2010}. The vacuum part of the model is done, first, in the BCRS, and then, converted to the GCRS moments of observation. Merely adding the geocentric atmospheric time delay to the vacuum part of the VLBI delay is unsatisfactory from theoretical point of view for we cannot be sure that all, perhaps minuscule but otherwise important relativistic contributions have been properly taken into account. Indeed, the Earth moves along with its atmosphere with respect to the BCRS, and radio signal propagates through the moving atmospheric medium in the barycentric coordinates. It introduces the specific Fresnel-Fizeau effect in the time of propagation of light through the atmosphere. This effect was not considered in the consensus VLBI model and it remained unclear what kind of the atmospheric time delay will remain after transformation of the Fresnel-Fizeau effect to the geocentric (GCRS) coordinates. The present paper gives an exhaustive analysis of this important theoretical problem.  

The rest of the paper is organized as follows.
Section \ref{sec2} gives a brief introduction to the consensus model. Section \ref{sec3} discusses the speed of light in a moving medium and Fresnel-Fizeau effect. Section \ref{sec4} derives the atmospheric time delay in the consensus model and proves that the Eubanks phenomenological model of the atmospheric delay and its coupling with relativity is correct. Discussion of our results is given in \ref{sec5}. 

We use $c$ to denote the constant fundamental speed in Minkowski space-time that is equal to the speed of light in vacuum. The locally-measured speed of light in medium is denoted as $c_\ell$. Three-dimensional spatial vectors are denoted with boldface letters, like ${\bm a}$, ${\bm b}$, etc.

\section{The Consensus VLBI Model} \la{sec2}
The relativistic consensus model for reduction of very long baseline (VLBI) observations was established about two decades ago \citep{eubanks_1991} and, currently, forms the basis of the IERS Standards \citep{IERS_2010}. The model is designed primarily for the analysis of geodetic VLBI observations of extragalactic objects (quasars) conducted from the surface of the Earth. In what follows, we shall consider specifically this case.

The basic equation of the consensus model yields the time delay, $t_2-t_1$, between the geocentric coordinate times (TCG) of arrivals of the front of a radio wave from a quasar to the first and second VLBI stations on the Earth's surface \cite[eq. 11.10]{IERS_2010}, 
\be\la{1} 
t_2-t_1=t_{\rm v_2}-t_{\rm v_1}+\lt(\delta t_{\rm atm_2}-\d t_{\rm atm_1}\ri)+\d t_{\rm atm_1}\frac{{\bm K}\cdot\lt({\bm w}_2-{\bm w}_1\ri)}{c}\;,
\ee
where, here and everywhere else, the dot between two vectors denote a usual Euclidean dot product, $t_{\rm v_2}-t_{\rm v_1}$ is the time delay of the radio wave as if it propagated in vacuum, ${\bm K}$ is the unit vector from the barycenter of the solar system to the source of light in the absence of gravitational or aberrational bending, $\d t_{\rm atm_i}$ are the atmospheric propagation delays measured in the local inertial frame at the sites of the first ($i=1$) and second ($i=2$) VLBI receivers, ${\bm w}_i=d{\bm x}_i/dt$ and ${\bm x}_i\equiv{\bm x}_i(t)$ are the GCRS velocity and radius vector of the $i$-th receiver respectively.

The vacuum delay is given by \cite[eq. 11.9]{IERS_2010} 
\ba\la{2}
t_{\rm v_2}-t_{\rm v_1}&=&-\frac{\displaystyle 1-\frac{(1+\g)U}{c^2}-\frac{{\bm V}_\oplus^2}{2c^2}-\frac{{\bm V}_\oplus\cdot{\bm w}_2}{c^2}}{\displaystyle 1+\frac{{\bm K}\cdot({\bm V}_\oplus+{\bm w}_2)}{c}}\frac{{\bm K}\cdot{\bm b}}{c}\\\nonumber
&&\phantom{+++++}-\frac{\displaystyle 1+\frac{{\bm K}\cdot{\bm V}_\oplus}{2c}}{\displaystyle 1+\frac{{\bm K}\cdot({\bm V}_\oplus+{\bm w}_2)}{c}}\frac{{\bm V}_\oplus\cdot{\bm b}}{c^2}\\\nonumber
&&\phantom{++++++++}+\frac{\Delta T_{\rm grav}}{\displaystyle 1+\frac{{\bm K}\cdot({\bm V}_\oplus+{\bm w}_2)}{c}}
\;,
\ea
where $\Delta T_{\rm grav}$ is the differential gravitational time delay \cite[eq. 11.7]{IERS_2010}, ${\bm V}_\oplus$ is the barycentric BCRS velocity of the geocenter, $U$ is the gravitational potential at the geocenter neglecting the effects of the Earth's mass, $\g$ is the PPN parameter equal to 1 in general relativity,  ${\bm b}\equiv{\bm x}_2(t_1)-{\bm x}_1(t_1)$ is the GCRS baseline vector at the time of arrival $t_1$.

The vacuum part of the relativistic time delay equation (\ref{2}) is well-established theoretically and its validity has been independently checked many times \citep{Kopeikin_2011_book}. The present paper is concerned with the calculation of the atmospheric time delay and its coupling with relativity that is the second and third terms in the right side of (\ref{1}). The current expression in (\ref{1})  was proposed by M. Eubanks who used the idea of the Lorentz transformation of the tropospheric delay between the two sites as measured in the proper time of the first observer. However, there is no exact theoretical justification/derivation of these terms in literature and it remained unclear why such particular form of the atmospheric time delay in (\ref{1}) was chosen. It is desirable to derive the atmospheric delay and its coupling with relativity from the first principles without relying upon plausible but in all other aspects phenomenological arguments given in support of these terms (see \cite[part III.e]{eubanks_1991}.

A standard approach for calculating the atmospheric time delay is non-relativistic \cite[Chapter 9]{IERS_2010}. It is based on assumption that the atmosphere is static with the speed of light, $c_{\ell}=c/n$, where $n$ is the refraction index of the atmosphere at the point under consideration. The refraction index $n$ depends on  pressure, temperature and humidity of the atmosphere as well as frequency of electromagnetic wave \cite[\S 7]{Murray_1983}. The pressure and temperature change with the altitude making the refraction index a function of spatial coordinates $n=n({\bm x})$. This approach does not take into account that calculation of the propagation of radio wave from quasar to VLBI receivers is done in the BCRS. Earth moves and rotates with respect to the BCRS, and so does the atmosphere. Hence, when one calculates the propagation of radio wave in BCRS, one has to take into account the relativistic effects caused by the moving refractive medium on the propagation of the radio wave. This gives rise to the Fresnel-Fizeau drag effect and introduces the  atmosphere-relativity coupling terms in VLBI time delay equation. Additional relativistic corrections arise from the transformation of the atmospheric time delay from the BCRS to observer's local frame \cite[Chapter 10]{IERS_2010}. The goal of the present paper is to account for the relativistic effects in the propagation of electromagnetic wave through the Earth's atmosphere and to derive the atmospheric VLBI time delay from the solid theoretical principles.

\section{Speed of Light in a Moving Medium and Fresnel-Fizeau effect.}\la{sec3}

In calculation of the coupling between the atmospheric time delay and relativistic terms it will be enough to take into account only special relativity as the gravity effects (e.g. the Shapiro delay, etc.) are small during the time of propagation of the signal in atmosphere and they can be neglected in the atmosphere-relativity coupling terms \footnote{The reader should be aware that the Shapiro time delay {\it in vacuum} due to the gravitational field of the Earth and calculated on the part of light trajectory lying in the atmosphere,
 is sufficiently large and cannot be neglected \citep{IERS_2010}}. Let us consider a small element of volume of the Earth atmosphere which is moving along with the Earth with respect to the BCRS, and rotates with the same angular velocity as the Earth. 

Let ${\bm u}$ be the BCRS velocity of motion of the volume element. Velocity of light in the comoving local frame of the volume element is ${\bm c}_{\ell}=c_\ell{\bm k}$, where $c_{\ell}=c/n$ is the speed of light in the rest frame of the element of the atmosphere, and ${\bm k}$ is the unit vector in the direction of propagation of the light ray. Velocity of light in the BCRS is given by the special relativistic law of the addition of velocities \cite[eq. 2.176]{Kopeikin_2011_book}
\be\la{2a}
{\bm c}_{L}=\frac{1}{\displaystyle 1+\frac{{\bm u}\cdot{\bm c}_{\ell}}{c^2}}\left[{\bm u}+\sqrt{1-{\frac{u^2}{c^2}}}{\bm c}_{\ell} +  \frac{1}{1+\sqrt{\displaystyle 1-\frac{u^2}{c^2}}}\frac{({\bm c}_{\ell} \cdot {\bm u}){\bm u}}{c^2}\right]\;,
\ee 
where ${\bm c}_L=c_L{\bm p}$, $c_L$ is the BCRS speed of light and ${\bm p}$ is the BCRS unit vector in the direction of propagation of light in the moving medium. We notice that the unit vector ${\bm p}\simeq -{\bm K}$ -- the difference in the directions of the two vectors is proportional to the discrepancy between $c$ and $c_\ell$ or, in other words, $n-1$. It comes also from the gravitational light-bending (and from the parallactic effects even if they are negligibly small in the case of quasars).
Equation (\ref{2a}) predicts that the BCRS speed of light, $c_L$, depends on the instantaneous value of the 3-velocity of 
motion of the element of the medium, ${\bm u}$.  

In what follows, it is sufficient to retain in (\ref{2a}) only the linear with respect to the velocity ${\bm u}$, terms. This is because the BCRS velocity, ${\bm u}$, is approximately equal to the orbital velocity of the Earth, ${\bm V}_\oplus$, which ratio to the fundamental speed is $V_\oplus/c\simeq 10^{-4}$, meaning that the quadratic with respect to velocity $\bm u$ corrections to the atmospheric time delay are too small to be measured.  Expanding numerator of (\ref{2a}) in a Taylor series with respect to the ratio, $u/c$, simplifies (\ref{2a}) to
\be\la{2b}
{\bm c}_{L}=\frac{{\bm c}_\ell+{\bm u}}{\displaystyle 1+\frac{{\bm u}\cdot{\bm c}_{\ell}}{c^2}}\;.
\ee
Calculating the absolute value of ${\bm c}_L$, and expanding the numerator of (\ref{2b}) with respect to $u/c$, yields the speed of light in the moving medium,
\be\la{2c}
c_L=c_\ell+\lt({\bm k}\cdot{\bm u}\ri)\lt(1-\frac{c_\ell^2}{c^2}\ri)\;,
\ee
where the residual terms are of the order of $(1-c^2_\ell/c^2)(u^2/c^2)$, and have been discarded.

The last term in the right side of (\ref{2c}) has been developed in 1830 by A. Fresnel, and confirmed in  1851 by A. Fizeau in his famous experiment with light travelling through moving water \citep{Stachel_2005}. Before the advent of special relativity, the effect was interpreted as a proof of the existence of partial ``aether dragging''. This explanation was abandoned in 1907 after Max von Laue \citep{Laue_1907} showed that special relativity predicts the result of the Fizeau experiment from the velocity addition theorem without any need for an aether.

\section{The Atmospheric Time Delay} \la{sec4}

The Earth atmosphere has two main parts - ionosphere and troposphere. The ionosphere contains charged particles - electrons, which make the ionospheric index of refraction, $n$ dependent on frequency of the electromagnetic signal. For this reason, the ionospheric time delay also depends on frequency. This property is used to calibrate the content of the electrons and to eliminate the ionospheric time delay from the VLBI time delay model. In what follows, we assume that it was done and the main contribution to the atmospheric time delay comes from the troposphere. Troposphere consists of neutral particles of air, and its index of refraction, $n$, can be considered to a large extent as frequency independent \cite[\S 7.4]{Murray_1983}.  

Theoretical calculation of the VLBI time delay begins from the consideration of propagation of light from the source of light (quasar) to observer in the barycentric coordinates $(T,{\bm X})$, where $T$ is the coordinate time (TCB) and ${\bm X}$ are the spatial BCRS coordinates. The light ray path consists of two parts - propagation in vacuum, and propagation through the Earth atmosphere. The fundamental VLBI equation in the BCRS can be written down in the following form \citep{Kopeikin_1990,Ashby_1991}
\be\la{3}
T_2-T_1=-\frac1c{\bm K}\cdot\lt({\bm X}_2-{\bm X}_1\ri)+\D T_{\rm grav}+\D T_{\rm atm};,
\ee
where ${\bm K}$ is the same unit vector as in (\ref{1}), $T_i$ is the barycentric time of arrival of the light signal to the $i$-th receiver, ${\bm X}_i$ are the BCRS coordinates of the receiver at the time $T_i$, $\D T_{\rm grav}$ is the same differential gravitational time delay as in (\ref{2}), and 
\be\la{4}
\D T_{\rm atm}=\d T_{\rm atm_2}-\d T_{\rm atm_1}\;,
\ee
is the BCRS differential delay due to the passage of the radio wave through the Earth atmosphere which moves along with the Earth and rotates. Relativistic transformations of (\ref{3}) to TCG time $t$ yields
\be\la{5}
t_2-t_1= t_{\rm v_2}-t_{\rm v_1}+\frac{\D T_{\rm atm}}{\displaystyle 1+\frac{{\bm K}\cdot({\bm V}_\oplus+{\bm w}_2)}{c}}\;,
\ee
where the vacuum time delay, $t_{\rm v_2}-t_{\rm v_1}$, is given in (\ref{2}).
Next step is to calculate $\d T_{\rm atm_i}$, and to express it in terms of the TCG time delay, that is $\d t_{\rm atm_i}$. Concluding this paragraph, we notice that the TCG time $t$ is a coordinate time scale defined in a region of space-time in the Earth's world tube neighborhood. It relates to the proper time $\tau_A$ measured along the world-line of an observer $A$ by an ordinary differential equation that can be found in IERS Standards \citep[Chapter 10, Eq. 10.6]{IERS_2010}.

The BCRS atmospheric time delay, $\d T_{\rm atm_i}$, calculated along the $i$-th light-ray path, is
\be\la{6}
\d T_{\rm atm_i}=\int\limits_{T_{ai}}^{T_i}\lt(\frac{c}{c_{L}}-1\ri)dT\;,
\ee
where where the integration is performed using $T$ as a parameter along the light ray, $c_{L}$ is the speed of light in the moving atmosphere taken from (\ref{2c}), $T_{ai}$ is the TCB time of entering the light ray to the atmosphere, $T_i$ is the TCB time of arrival of the light signal to the $i$-th receiver. 

Equation (\ref{2c}) gives 
\be\la{7}
\frac{c}{c_{L}}=n-\frac1c\lt({\bm k}\cdot{\bm u}\ri)\lt(n^2-1\ri)\;,
\ee
where the index of refraction, $n$, differs from 1 by a very small amount, $n-1\simeq 3\times 10^{-4}$ at normal atmospheric pressure and room temperature \cite[\S 7.4]{Murray_1983}. Hence, with a sufficient approximation, $n^2-1=2(n-1)$, and the integral (\ref{6}) is reduced to
\be\la{8}
\d T_{\rm atm_i}=\int\limits_{T_{ai}}^{T_i}\lt(n-1\ri)\lt(1-\frac2c{\bm k}\cdot{\bm u}\ri)dT\;.
\ee
where all quantities in the integrand are taken on the path of the $i$-th light ray.
\begin{figure}[h]
\includegraphics[width=20pc]{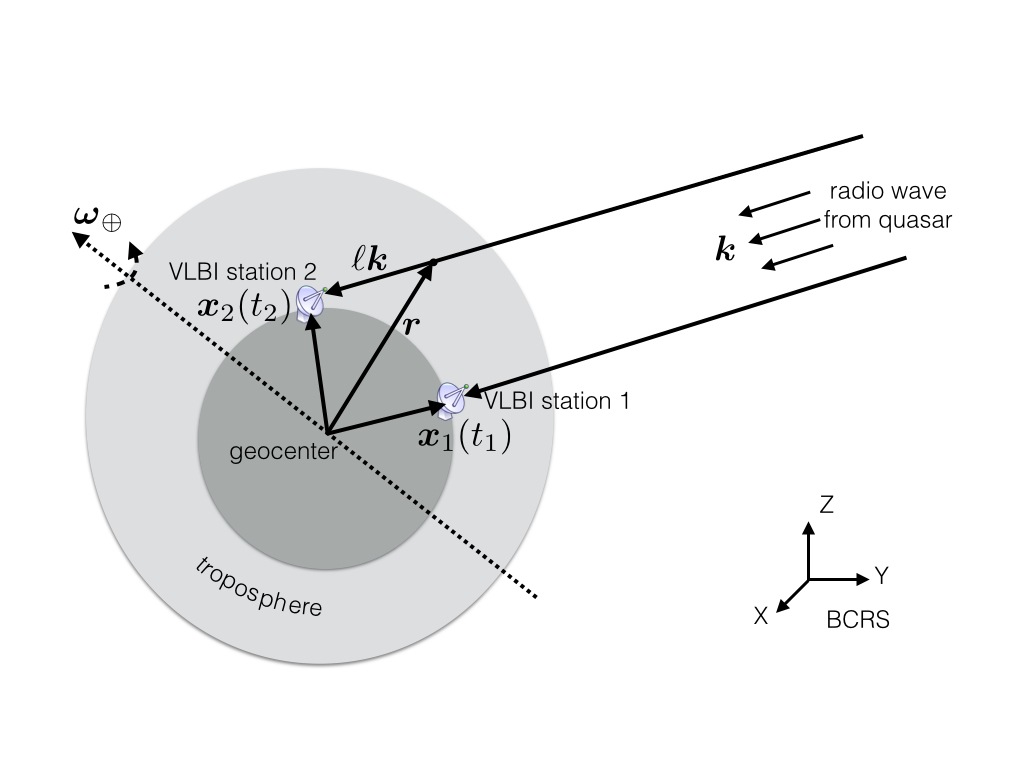}
\caption{Geometry of VLBI observation. A plane radio wave from quasar travels in the direction shown by the unit vector ${\bm k}$. Vectors ${\bm x}_1$ and ${\bm x}_2$ point out positions of VLBI receivers at the time of arrival of the plane front of a radio wave from quasar to each radio dish. Vector ${\bm r}$ point out to the position of the radio signal at time $t$ and is defined in (\ref{10}). Vector ${\bm\omega}_\oplus$ is the angular velocity of Earth's diurnal rotation.}
\label{figone}
\end{figure}

BCRS velocity, ${\bm u}$, of the element of troposphere can be decomposed in a linear sum of two terms
\be\la{9}
{\bm u}={\bm V}_\oplus+{\bm w}_{\rm atm}\;,
\ee
where ${\bm V}_\oplus$ is the BCRS velocity of the geocenter, and ${\bm w}_{\rm atm}$ is velocity of the atmosphere on the light ray trajectory with respect to the geocentric celestial reference system (GCRS). Equation (\ref{9}) neglects all relativistic terms of the order of $1/c^2$ which are too small for practical measurement in the atmospheric time delay model (but they might be interesting for more deep theoretical study of the relativistic propagation of light in moving media in some astrophysical situations like rapidly rotating pulsar's magnetosphere).
The troposphere follows rotation of the Earth around its axis with the angular velocity, ${\bm\omega}$, so that we can write
\be\la{10}
{\bm w}_{\rm atm}= {\bm\omega}_\oplus\times{\bm r}\;,
\ee
where ${\bm r}$ is the GCRS radius vector of quasar's radio signal in atmosphere. It can be presented as an algebraic sum of two radius vectors (see Fig. \ref{figone}),
\be\la{11}
{\bm r} = {\bm x}_i-\ell{\bm k}\;,
\ee
where ${\bm x}_i$ is the GCRS radius vector of the $i$-th VLBI receiver, and $\ell$ is the geocentric spatial distance between the receiver and the current position of the radio signal. 

We substitute (\ref{10}), (\ref{11}) to (\ref{9}), and notice that ${\bm k}\cdot\lt({\bm\omega}_\oplus\times{\bm k}\ri)=0$, while ${\bm\omega}_\oplus\times{\bm x}_i={\bm w}_i$ is the GCRS velocity of the $i$-th receiver. Then, equation (\ref{8}) takes on the following form,
\be\la{12}
\d T_{\rm atm_i}=\d {\cal T}_{\rm atm_i}\lt[1-\frac2c{\bm k}\cdot\lt({\bm V}_\oplus+{\bm w}_i\ri)\ri]\;,
\ee
where 
\be\la{13}
\d {\cal T}_{\rm atm_i}\equiv \int\limits_{T_{ai}}^{T_i}\lt(n-1\ri)dT\;,
\ee
is the TCB interval of time corresponding to the time delay calculated in the static (non-moving) atmosphere in the BCRS coordinates. This time interval is not directly measurable, and should be transformed to the time delay, $\d t_{\rm atm_i}$, calculated in the local inertial frame of observer located at the site of the $i$-th receiver. It suffices to account for only special relativistic terms in this transformation which yields
\be\la{14}
\d {\cal T}_{\rm atm_i}=\d t_{\rm atm_i}\lt[1+\frac1c{\bm k}\cdot\lt({\bm V}_\oplus+{\bm w}_i\ri)\ri]\;,
\ee
where we have taken into account that the time transformation has been performed between the two events taken on the light ray.

Equation (\ref{14}) is plugged into (\ref{13}) and the unit vector ${\bm k}$ is replaced with the unit vector ${\bm K}=-{\bm k}$ directed from the barycenter of the solar system to the source of light. It yields,
\be\la{15}
\d T_{\rm atm_i}=\d t_{\rm atm_i}\lt[1+\frac1c{\bm K}\cdot\lt({\bm V}_\oplus+{\bm w}_i\ri)\ri]\;,
\ee
This expression is used in (\ref{4}) to calulate $\D T_{\rm atm}$ that enters numerator of the second term in the right side of (\ref{5}). It results in
the Eubanks phenomenological equation (\ref{1}) thus, giving a rigorous theoretical justification for its appearance in the VLBI consensus model.

\section{Discussion}\la{sec5}

We have presented an exact theoretical derivation of the atmospheric time delay in the consensus model with an accounting for the coupling with (special) relativistic terms. We have shown that the calculation of the atmospheric time delay requires a proper discussion of the Fresnel-Fizeau drag effect (\ref{2c}) in the BCRS and relativistic time transformation formula (\ref{14}) of time intervals between two events (the time of entering atmosphere and the time of observation) taken on the light ray propagating in atmosphere.

We have found that the criticism of the tropospheric time delay equations given in the works by B. Shahid-Saless et al. \citep{Ashby_1991} and by I. Shapiro (1983, unpublished) given in \citep[section III.e]{eubanks_1991} is unfounded. In fact, both Shahid-Saless and Shapiro referred their tropospheric time delay to the BCRS, which is equivalent to our formula (\ref{5}). Taking into account the Fresnel-Fizeau effect and the special relativistic time transformation reduces (\ref{5}) exactly to the expression (\ref{1}) given in \citep{eubanks_1991}.

SLR/LLR analysis is done in the inertial (non-rotating) GCRS. Rotational motion of atmosphere causes the Fresnel-Fizeau effect in one-way light propagation model. However, in SLR/LLR experiments light propagates troposphere twice - from the ground station to retro-reflector and back. The unit vector, ${\bm k}$, entering the Fresnel-Fizeau effect (\ref{2c}) reverses direction while velocity, ${\bm u}$ of the troposphere remains practically unchanged during the travel time of the light pulse. For this reason, the excess time delay caused by the Fresnel-Fizeau effect cancels out in SLR/LLR technique which corresponds to the currently used light propagation model. Thus, we conclude that the atmospheric time delay and/or its coupling with relativity in the SLR/LLR/VLBI geodetic techniques cannot be responsible for the observed scaling offset \citep{Altamimi_2011} between the two versions of ITRF based independently on VLBI and SLR techniques. 

Finally, we emphasize that the present paper was concerned with the derivation of the Fresnel-Fizeau effect and its appearance in the consensus model in the linear, with respect to velocity of the medium, terms. It is rather straightforward to apply our formalism for calculation the quadratic, and higher-order terms, if necessary. In particular, the linear Fresnel-Fizeau effect associated with the velocity ${\bm V}_\oplus$ of the orbital motion of the Earth does not appear in the final form of the atmospheric time delay equation (\ref{1}). It  might be interesting to check if the quadratic Fresnel-Fizeau effect associated with the velocity ${\bm V}_\oplus$ of the orbital motion of the Earth, would appear in the atmospheric time delay equation. Such a dependence is unlikely from relativistic point of view as it contradicts to the principle of relativity \citep{Kopeikin_2011_book} but its mathematical proof is still required.
\begin{acknowledgements} 
We are grateful to M. Eubanks, J.~C. Ries, G. Petit and E. Pavlis for useful conversations and to J.~C. Ries for providing us with the Power Point slides of his talk given at IERS Unified Analysis Workshop, Pasadena, CA (27-28 June 2014). We are also thankful to two anonymous referees for fruitful comments and suggestions for improving the manuscript as well as to Prof. Dr. Johannes B\"ohm for his kind editorial guidance through the publication process. 

Data to support this article are from the International Earth Rotation and Reference Systems Service as well as from the International Laser Ranging Service of the International Association of Geodesy.

S.~M. Kopeikin has been supported by the Faculty Fellowship 2014 in the College of Arts and Science of the University of Missouri and the grant \textnumero 14-27-00068 of the Russian Scientific Foundation. The work of W.-B. Han has been funded by the National Natural Science Foundation of China (grant \textnumero 11273045) and the China Scholarship Council Fellowship \textnumero 201304910030.
\end{acknowledgements}

\bibliographystyle{apalike}
\bibliography{atmospheric_delay}
\end{document}